\documentclass[11pt,onecolumn,draftclsnofoot]{IEEEtran}

\usepackage[left=1.25in, right=1.25in, bottom=1in, top=1in]{geometry}
\usepackage[english]{babel}

\usepackage{amsmath}
\usepackage{amssymb}
\usepackage{amsthm}
\usepackage{amsfonts}
\usepackage{mathtools}
\usepackage[binary-units=true]{siunitx}
\newcommand{\mat}[1]{\ensuremath{{\mathbf{\MakeUppercase{#1}}}}}
\makeatletter
\renewcommand{\vec}[1]{%
	\ifcat\relax\noexpand#1%
	\ensuremath{\boldsymbol{\lowercase{#1}}}%
	\else
	\ensuremath{\mathbf{\lowercase{#1}}}%
	\fi
}
\makeatother
\newcommand{\ev}[1]{\ensuremath{\mathbb{E}\left\{ #1 \right\}}}
\newcommand{\transpose}[1]{\ensuremath{{#1}^{\textsc{t}}}}
\newcommand{\inverse}[1]{\ensuremath{{#1}^{-1}}}
\newcommand{\R}{\ensuremath{\mathbb{R}}}
\newcommand{\norm}[1]{\left|\left|#1\right|\right|}
\newcommand{\sumlim}[2]{\ensuremath{\sum\limits_{#1}^{#2}}}

\usepackage{caption}
\usepackage{subcaption}
\usepackage{graphicx}
\usepackage{xcolor}

\usepackage{booktabs}
\usepackage{tabularx}
\usepackage{verbatimbox}

\usepackage{tikz}
\usetikzlibrary{external,automata,trees,shadows,matrix,fit,decorations,positioning,shapes,plotmarks,patterns,arrows,calc,quotes,tikzmark,arrows.meta,plotmarks,decorations.pathreplacing,backgrounds}

\definecolor{blue-sg}{rgb}{0.00000,0.44700,0.74100}

\usepackage[most]{tcolorbox}
\usepackage[noadjust]{cite}
\usepackage{hyperref}
\makeatletter
\newcommand{\gettikzxy}[3]{%
	\tikz@scan@one@point\pgfutil@firstofone#1\relax
	\edef#2{\the\pgf@x}%
	\edef#3{\the\pgf@y}%
}
\makeatother

\title{EEG-based Auditory Attention Decoding\\ \LARGE Towards Neuro-Steered Hearing Devices}
\author{Simon~Geirnaert, Servaas~Vandecappelle, Emina~Alickovic, Alain~de~Cheveign\'e, Edmund~Lalor, Bernd~T.~Meyer, Sina~Miran, Tom~Francart, and~Alexander~Bertrand
	\thanks{\noindent This research is funded by an Aspirant Grant from the Research Foundation - Flanders (FWO) (for S. Geirnaert, nr. 1136219N), the KU Leuven Special Research Fund C14/16/057, FWO project nr. G0A4918N, the European Research Council (ERC) under the European Union’s Horizon 2020 research and innovation programme (grant agreement No 802895 and grant agreement No 637424), and the Flemish Government under the “Onderzoeksprogramma Artificiële Intelligentie (AI) Vlaanderen” programme. The scientific responsibility is assumed by its authors.}
	\thanks{\noindent The first two authors have implemented all the algorithms of the comparative study to ensure uniformity. All implementations have been checked and approved by at least one of the authors of the original paper in which the method was presented.}
}

\begin{document}
	
	\maketitle
	
	\vspace{-1.5cm}
	
	\begin{abstract}
		People suffering from hearing impairment often have difficulties participating in conversations in so-called `cocktail party' scenarios with multiple people talking simultaneously. Although advanced algorithms exist to suppress background noise in these situations, a hearing device also needs information on which of these speakers the user actually aims to attend to. The correct (attended) speaker can then be enhanced using this information, and all other speakers can be treated as background noise. Recent neuroscientific advances have shown that it is possible to determine the focus of auditory attention from non-invasive neurorecording techniques, such as electroencephalography (EEG). Based on these new insights, a multitude of auditory attention decoding (AAD) algorithms have been proposed, which could, combined with the appropriate speaker separation algorithms and miniaturized EEG sensor devices, lead to so-called neuro-steered hearing devices. In this paper, we provide a broad review and a statistically grounded comparative study of EEG-based AAD algorithms and address the main signal processing challenges in this field.    
	\end{abstract}
	
	\section{Introduction}
	\label{sec:introduction}
	\noindent
	Current state-of-the-art hearing devices, such as hearing aids or cochlear implants, contain advanced signal processing algorithms to suppress acoustic background noise and, as such, assist the constantly expanding group of people suffering from hearing impairment. However, situations where multiple competing speakers are active simultaneously (dubbed the `cocktail party problem') still cause major difficulties for the hearing device user, often leading to social isolation and decreased quality of life. Beamforming algorithms that use microphone array signals to suppress acoustic background noise and extract a single speaker from a mixture lack a fundamental piece of information to assist the hearing device user in cocktail party scenarios: which speaker should be treated as the attended speaker (i.e., the speaker to which the user/listener intends to attend to) and which other speaker(s) should be treated as the interfering noise source(s)? This issue is often addressed using simple heuristics, for example, by selecting the loudest speaker or assuming that the attended speaker is in front of the listener. However, in many practical situations, these heuristics will select and enhance a speaker that is not attended by the user. For example, when listening to a passenger while driving a car or when listening to a public address system, a selection based on the look direction will fail.
	
	Recent neuroscientific insights on how the brain synchronizes with the speech envelope~\cite{mesgarani2012selective,ding2012emergence} have laid the groundwork for a new strategy to tackle this problem: extracting attention-related information directly from the origin, i.e., the brain. This is generally referred to as the ‘auditory attention decoding’ (AAD) problem. In the last ten years, following these groundbreaking advances in the field of auditory neuroscience and neural engineering, the topic of AAD has gained traction in the biomedical signal processing community. AAD can be performed based on several neurorecording modalities, such as electroencephalography (EEG)~\cite{o2014attentional}, electrocorticography (ECoG)~\cite{mesgarani2012selective} or magnetoencephalography (MEG)~\cite{ding2012emergence}. However, the invasiveness of ECoG and the high cost and lack of wearability of MEG limit their applicability in practical hearing devices for daily-life usage. On the other hand, EEG is considered to be a good candidate to be integrated with hearing devices as it is a non-invasive, wearable, and relatively cheap neurorecording technique. 
	
	In~\cite{o2014attentional}, a first successful speech-based AAD algorithm based on unaveraged single-trial EEG data was proposed. The main idea of~\cite{o2014attentional} is to decode the attended speech envelope from a multi-channel EEG recording using a neural decoder and correlate the decoder output with the speech envelope of each speaker. Following this seminal work, many new AAD algorithms have been developed~\cite{biesmans2017auditory,Alick2019,miran2018real,de2018decoding,de2017machine,Ciccarelli2019,Vandecappelle475673}. In combination with effective speaker separation algorithms~\cite{van2017eeg,o2017neural,Han2019,pu2019joint,aroudi2020cognitive} and relying on rapidly evolving improvements in miniaturization and wearability of EEG sensors~\cite{Kappel2019,Debener2015,mirkovic2016target,narayanan2019analysis}, these advances could lead to a new assistive solution for the hearing impaired: a \emph{neuro-steered hearing device}.
	
\begin{figure}
		\centering
		\includegraphics[width=0.9\linewidth]{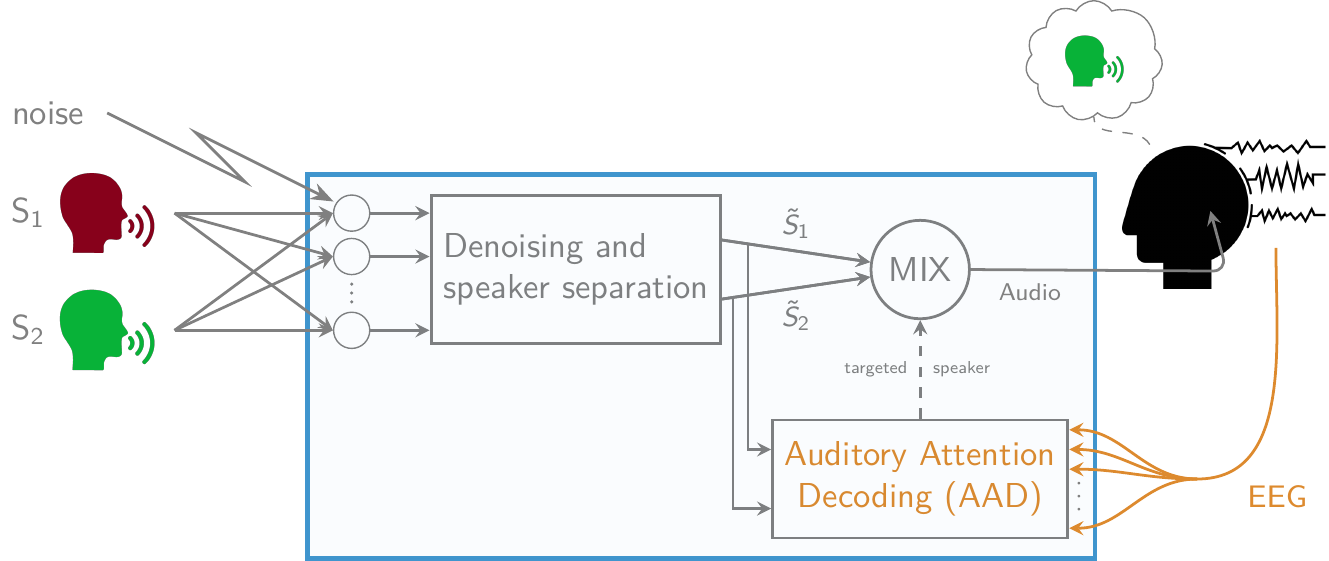}
		\caption{A conceptual overview of a neuro-steered hearing device when there are two competing speakers. The green speaker (S\textsubscript{2}) corresponds to the attended one, while the red speaker (S\textsubscript{1}) corresponds to the unattended one.}
		\label{fig:neuro-steered-HA}
	\end{figure}

	Fig.~\ref{fig:neuro-steered-HA} shows a conceptual overview of a neuro-steered hearing device when there are two competing speakers. The AAD block contains an algorithm that determines the attended speaker by integrating the demixed speech envelopes and the EEG. Despite the large variety in AAD algorithms, an objective and transparent comparative study has not been performed to date, making it hard to identify which strategies are most successful. In this paper, we will briefly review different types of AAD algorithms and their most common instances, and provide an objective and quantitative comparative study using two independent, publicly available datasets~\cite{das_neetha_2019_3377911,fuglsang_soren_a_2018_1199011}. This comparative study has been reviewed and endorsed by the author(s) of the original papers in which these algorithms were proposed to ensure fairness and correctness. While the paper's main focus is on this AAD block, we also provide an outlook on other practical challenges on the road ahead, such as the evaluation in more realistic listening scenarios, the interaction of AAD with speech demixing or beamforming algorithms, and challenges related to EEG sensor miniaturization.

	\section{Review of AAD algorithms}
	\label{sec:review}
	\noindent
	In this section, we provide a brief overview of various AAD algorithms. This comparative study includes only papers published before the year 2020, when this paper was conceptualized. However, since this field is progressing fast and several new papers have appeared since the conceptualization of this article, the reader is encouraged to look up new AAD algorithms (and extensions thereof) and compare them with the presented methods.
	
	For the sake of an easy exposition, we assume that there are only two speakers (one attended and one unattended speaker), although all algorithms can be generalized to more than two speakers. In the remainder of this paper, we also make abstraction of the speaker separation and denoising block in Fig.~\ref{fig:neuro-steered-HA} and assume that the AAD block has direct access to the envelopes of the original unmixed speech sources as often done in the AAD literature. However, we will briefly return to the combination of both blocks in Section~\ref{sec:sp-sep-eeg-minia}. 
	
	Most AAD algorithms adopt a \emph{stimulus reconstruction} approach (also known as backward modeling or decoding). In this strategy, a multi-input single-output (MISO) neural decoder is applied to all EEG channels to reconstruct the attended speech envelope. This neural decoder is pre-trained to optimally reconstruct the attended speech envelope from the EEG data while blocking other (unrelated) neural activity. It is in this training procedure that most AAD algorithms differ. The reconstructed speech envelope is afterwards correlated with the speech envelopes of all speakers, after which the one with the highest Pearson correlation coefficient is identified as the attended speaker (Fig.~\ref{fig:overview-stim-rec-nn-sr}). This correlation coefficient is estimated over a window of $\tau$ seconds, which is referred to as the \emph{decision window length}, corresponding to the amount of EEG data used in each decision on the attention. Typically, the AAD accuracy strongly depends on this decision window length because the Pearson correlation estimates are very noisy due to the low signal-to-noise ratio of the output signal of the neural decoder.
	
	Alternatively, the neural response in each EEG channel can be predicted from the speech envelopes via an encoder (also known as forward modeling or encoding) and can then be correlated with the measured EEG~\cite{Wong2018,Alick2019}. When the encoder is linear, this corresponds to estimating impulse responses (aka temporal response functions) between the speech envelope(s) and the recorded EEG signals. For AAD, backward MISO decoding models have been demonstrated to outperform forward encoding models~\cite{Wong2018,Alick2019}, as the former can exploit the spatial coherence across the different EEG channels at its input. In this comparative study, we thus only focus on backward AAD models, except for the canonical correlation analysis (CCA) algorithm (Section~\ref{sec:cca}), which combines both a forward and backward approach.
	
	Due to the emergence of deep learning methods, a third approach has become popular: \emph{direct classification}~\cite{Vandecappelle475673,Ciccarelli2019}. In this approach, the attention is directly predicted in an end-to-end fashion, without explicitly reconstructing the speech envelope.
	
	The decoder models are typically trained in a supervised fashion, which means that the attended speaker must be known for each data point in the training set. This requires collecting `ground-truth' EEG data during a dedicated experiment in which the subject is asked to pay attention to a predefined speaker in a speech mixture. The models can be trained either in a \textit{subject-specific} fashion (based on EEG data from the actual subject under test) or in a \textit{subject-independent} fashion (based on EEG data from other subjects than the subject under test). The latter leads to a universal (subject-independent) decoder, which has the advantage that it can be applied to new subjects without the need to go through such a tedious ground-truth EEG data collection for every new subject. However, since each person's brain responses are different, the accuracy achieved by such universal decoders is typically lower~\cite{o2014attentional}. In this paper, we only consider subject-specific decoders, which allows achieving better accuracies, as they are tailored to the EEG of the specific end-user. Transfer learning techniques, which are becoming popular in the field of brain-computer interfaces~\cite{lotte2017review}, may close the gap between subject-specific and subject-independent models, although this remains to be researched in the context of AAD.
	
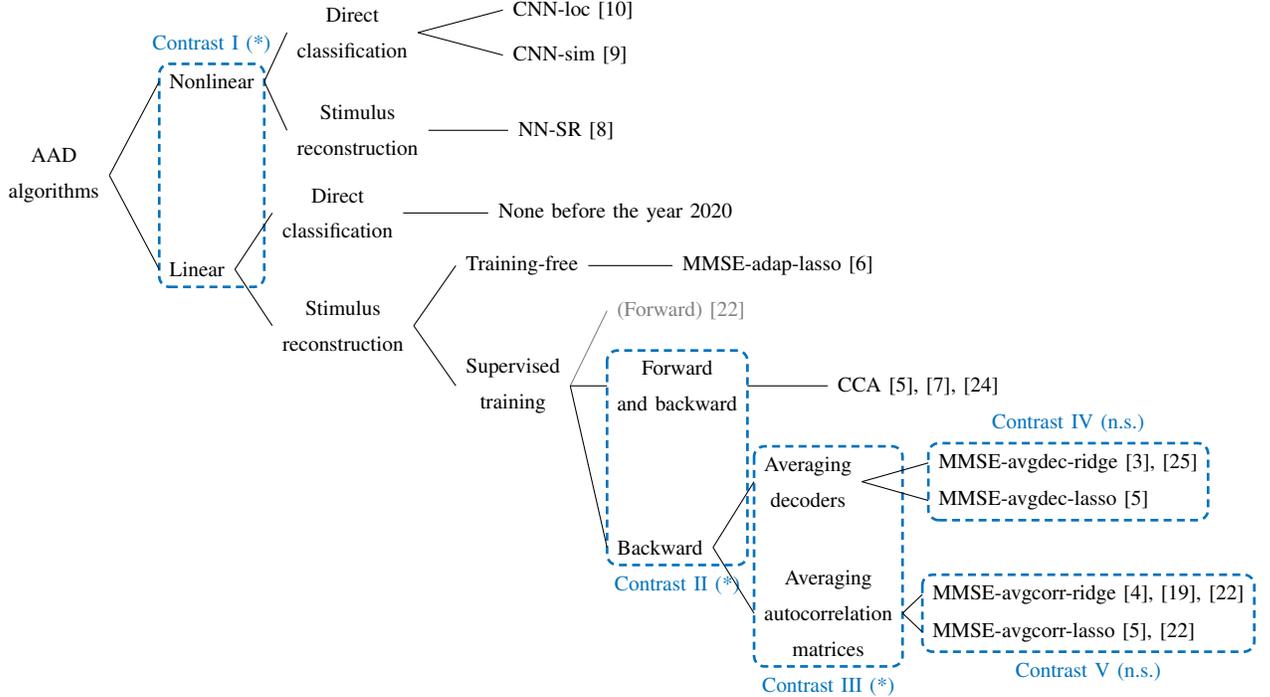
\begin{figure*}
		\centering
		\scalebox{1}{
			\begin{tikzpicture}[level distance=2cm,
				level 1/.style={sibling distance=2.5cm},
				level 2/.style={sibling distance=2cm},
				level 3/.style={sibling distance=1.6cm},
				level 4/.style={sibling distance=1.75cm},
				level 5/.style={sibling distance=1.75cm},
				level 6/.style={sibling distance=0.5cm},
				font=\scriptsize,grow'=right,child anchor=west, parent anchor=east]
				\node[align=center] (tree) {AAD \\algorithms}
				child {node[anchor=west,xshift=-0.6cm] (nl) {Nonlinear}
					child {node[anchor=west,yshift=-0.35cm,xshift=-1cm,align=center] (dc) {Direct \\classification}
						child {node[anchor=west,yshift=-0.5cm] (cnnd) {CNN-loc~\cite{Vandecappelle475673}}}
						child {node[anchor=west,yshift=0.5cm] (cnnc) {CNN-sim~\cite{Ciccarelli2019}}}}
					child {node[anchor=west,yshift=0.35cm,xshift=-1cm,align=center] (sr) {Stimulus \\ reconstruction}
						child {node[anchor=west] {NN-SR~\cite{de2017machine}}}
				}}
				child {node[anchor=west,xshift=-0.6cm] (lin) {Linear}
					child {node[anchor=west,xshift=-1cm,yshift=-0.25cm,align=center] {Direct \\classification}
						child {node[anchor=west,xshift=-0cm] {None before the year 2020}}}
					child {node[anchor=west,xshift=-1cm,yshift=0.25cm,align=center] {Stimulus \\reconstruction}
						child {node[anchor=west,xshift=-0.5cm] {Training-free}
							child {node[anchor=west] {MMSE-adap-lasso~\cite{miran2018real}}}
						}
						child {node[anchor=west,xshift=-0.5cm,align=center] {Supervised \\ training}
							child[draw=gray] {node[gray,anchor=west,align=center,xshift=-0.75cm,yshift=-0.75cm] (fwb) {(Forward)~\cite{Wong2018}}}
							child {node[anchor=west,align=center,xshift=-0.75cm] (fwb) {Forward \\and backward}
								child {node[anchor=west,align=center] {CCA~\cite{de2018decoding,Dmochowski2018,Alick2019}}}
							}
							child {node[anchor=west,xshift=-0.75cm,yshift=-0.4cm] (bw) {Backward}
								child {node[anchor=west,align=center,xshift=-0.75cm] (avgdec) {Averaging \\decoders}
									child {node[anchor=west,xshift=-0.4cm] (avgdec-ridge) {MMSE-avgdec-{ridge}~\cite{o2014attentional,schafer2018testing}}}
									child {node[anchor=west,xshift=-0.4cm] (avgdec-lasso) {MMSE-avgdec-{lasso}~\cite{Alick2019}}}
								}
								child {node[anchor=west,align=center,xshift=-0.75cm] (avgcorr) {Averaging \\autocorrelation \\matrices}
									child {node[anchor=west,xshift=-0.75cm] (avgcorr-ridge) {MMSE-avgcorr-{ridge}~\cite{biesmans2017auditory,Wong2018,narayanan2019analysis}}}
									child {node[anchor=west,xshift=-0.75cm] (avgcorr-lasso) {MMSE-avgcorr-{lasso}~\cite{Alick2019,Wong2018}}}
								}
							}
						}
					}
				};
				\scoped[on background layer]{
					\gettikzxy{(lin.south east)}{\fx}{\fy}
					\gettikzxy{(nl.north east)}{\sx}{\sy}
					\draw[blue-sg,line width=1,rounded corners,densely dashed] (lin.south west) -- (\sx,\fy) -- (nl.north east) to node[above,blue-sg] {Contrast I (*)} (nl.north west) -- cycle;
					\gettikzxy{(bw.south east)}{\fx}{\fy}
					\gettikzxy{(fwb.north east)}{\sx}{\sy}
					\draw[blue-sg,line width=1,rounded corners,densely dashed] (fwb.north west) -- (bw.south west) to node[below,blue-sg] {Contrast II (*)} (\sx,\fy) -- (fwb.north east) -- cycle;
					\gettikzxy{(avgcorr.south east)}{\fx}{\fy}
					\gettikzxy{(avgdec.north east)}{\sx}{\sy}
					\draw[blue-sg,line width=1,rounded corners,densely dashed] (avgdec.north west) -- (avgcorr.south west) to node[below,blue-sg] {Contrast III (*)} (avgcorr.south east) -- (\fx,\sy) -- cycle;
					\gettikzxy{(avgdec-lasso.south east)}{\fx}{\fy}
					\gettikzxy{(avgdec-ridge.north east)}{\sx}{\sy}
					\draw[blue-sg,line width=1,rounded corners,densely dashed] (avgdec-lasso.south west) -- (\sx,\fy) -- (avgdec-ridge.north east) to node[above,blue-sg] {Contrast IV (n.s.)} (avgdec-ridge.north west) -- cycle;
					\gettikzxy{(avgcorr-lasso.south east)}{\fx}{\fy}
					\gettikzxy{(avgcorr-ridge.north east)}{\sx}{\sy}
					\draw[blue-sg,line width=1,rounded corners,densely dashed] (avgcorr-ridge.north west) -- (avgcorr-lasso.south west) to node[below,blue-sg] {Contrast V (n.s.)} (\sx,\fy) -- (avgcorr-ridge.north east) -- cycle;
				}
		\end{tikzpicture}}
		\caption{The included AAD algorithms in this comparative study (except for the forward models; see the introduction of Section~\ref{sec:review}) and the planned contrasts in the statistical analysis. (*) indicates a significant difference ($p<0.05$), while (n.s.) indicates a non-significant difference (see Section~\ref{sec:stat-analysis} for more details).}
		\label{fig:tree-algos}
	\end{figure*}

	Fig.~\ref{fig:tree-algos} depicts a complete overview and classification of all algorithms included in our comparative study, discriminated based on their fundamental properties. In the following sections, we distinguish between linear and nonlinear algorithms.
	
	\subsection{Linear methods}
	\label{sec:linear-methods}
	\noindent
	All linear methods included in this study, which differ in the features shown in the linear branch of Fig.~\ref{fig:tree-algos}, adopt the so-called stimulus reconstruction framework (Fig.~\ref{fig:overview-stim-rec-nn-sr}). This boils down to applying a linear time-invariant spatio-temporal filter $d_c(l)$ on the $C$-channel EEG $x_c(t)$ to reconstruct the attended speech envelope $s_{\textnormal{a}}(t)$:
	\begin{equation}
		\label{eq:spatio-temporal-filter}
		\hat{s}_{\textnormal{a}}(t) = \sumlim{c = 1}{C}\sumlim{l = 0}{L-1}d_c(l)x_c(t+l),
	\end{equation}
	where $c$ is the channel index, ranging from $1$ to $C$, and $l$ is the time lag index, ranging from $0$ to $L-1$ with $L$ the per-channel filter length. The corresponding MISO filter is anti-causal, as the brain responds to the stimulus, such that only future EEG time samples can be used to predict the current stimulus sample. Eq. \eqref{eq:spatio-temporal-filter} can be rewritten as $\hat{s}_{\textnormal{a}}(t) = \transpose{\vec{d}}\vec{x}(t)$, using $\vec{d} \in \R^{LC \times 1}$, collecting all decoder coefficients for all time lags and channels, and $\vec{x}(t) = \transpose{\begin{bmatrix} \transpose{\underline{\vec{x}}_1(t)} & \transpose{\underline{\vec{x}}_2(t)} & \cdots & \transpose{\underline{\vec{x}}_C(t)} \end{bmatrix}} \in \R^{LC \times 1}$, with $\underline{\vec{x}}_c(t) = \transpose{\begin{bmatrix} x_c(t) & x_c(t+1) & \cdots & x_c(t+L-1) \end{bmatrix}}$ (the same indexing holds for the decoder $\vec{d}$).
	
	In the next three sections, we introduce the different linear methods included in this study. These linear methods, which are all correlation-based, can be extended to more than two competing speakers by simply correlating the reconstructed speech envelope with all additional speech envelopes of the individual competing speakers and taking the maximum.
	
	\subsubsection{Supervised minimum mean-squared error backward modeling (MMSE)}
	\label{sec:mmse-bw}
	\noindent
	The most basic way of training the decoder, first presented in the EEG-based AAD-context in~\cite{o2014attentional}, is by minimizing the mean-square error (MSE) between the actual attended envelope and the reconstructed envelope. In~\cite{biesmans2017auditory}, it is shown that minimizing the MSE is equivalent to maximizing the Pearson correlation coefficient between the reconstructed and attended speech envelope. Using sample estimates, assuming that there are $T$ samples available, the MMSE-based formulation becomes equivalent to the least-squares (LS) formulation:
	\begin{equation}
		\label{eq:ls}
		\hat{\vec{d}} = \underset{\vec{d}}{\textnormal{argmin}} \norm{\vec{s}_{\textnormal{a}}-\mat{X}\vec{d}}_2^2,
	\end{equation}
	with $\mat{X} = \transpose{\begin{bmatrix} \vec{x}(0) &\cdots& \vec{x}(T-1) \end{bmatrix}} \in \R^{T \times LC}$ and $\vec{s}_{\textnormal{a}} = \transpose{\begin{bmatrix} s_{\textnormal{a}}(0) &\cdots& s_{\textnormal{a}}(T-1) \end{bmatrix}} \in \R^{T \times 1}$. The normal equations lead to the solution $\hat{\vec{d}} = \inverse{\left( \transpose{\mat{X}}\mat{X}\right)} \transpose{\mat{X}}\vec{s}_{\textnormal{a}}$. The first factor corresponds to an estimation of the autocorrelation matrix $\hat{\mat{R}}_{xx} = \frac{1}{T} \sumlim{t = 0}{T-1} \vec{x}(t) \transpose{\vec{x}(t)} \in \R^{LC\times LC}$, while the second factor corresponds to the cross-correlation vector $\hat{\vec{r}}_{xs_{\textnormal{a}}} = \frac{1}{T} \sumlim{t = 0}{T-1} \vec{x}(t) s_{\textnormal{a}}(t) \in \R^{LC \times 1}$.
	
	To avoid overfitting, two types of regularization are used in AAD literature: ridge regression/L\textsubscript{$2$}-norm regularization and L\textsubscript{$1$}-norm/sparse regularization, also known as the least absolute shrinkage and selection operator (lasso). The corresponding cost functions are shown in Table~\ref{tab:mmse-bw}, where the regularization hyperparameter $\lambda$ is defined relative to $z = \frac{\textnormal{trace}(\transpose{\mat{X}}\mat{X})}{LC}$ (for ridge regression)/$q = \norm{\transpose{\mat{X}}\vec{s}_{\textnormal{a}}}_\infty$ (for lasso). Similar to~\cite{Alick2019}, we here use the alternating direction method of multipliers (ADMM) to iteratively obtain the solution of the lasso problem. The optimal value $\lambda$ can be found using a cross-validation scheme. Other regularization methods, such as Tikhonov regularization, have been proposed as well~\cite{Wong2018}.
	
	Assume a given training set consisting of $K$ data segments of a specific length $T$. These segments can either be constructed artificially by segmenting a continuous recording (usually for the sake of cross-validation), or they can correspond to different experimental trials (potentially from different subjects, e.g., when training a subject-independent decoder). There exist various flavors of combining these different segments in the process of training a decoder. As suggested in the seminal paper of~\cite{o2014attentional}, decoders $\mathbf{d}_k$ can be trained per segment $k$, after which all decoders are averaged to obtain a single, final decoder $\mathbf{d}$. In~\cite{biesmans2017auditory} (also adopted in, e.g., \cite{das2016effect,van2017eeg,aroudi2019impact,narayanan2019analysis,geirnaert2020interpretable,aroudi2020cognitive}), an alternative scheme is proposed, where, instead of estimating a decoder per segment separately, the loss function~\eqref{eq:ls} (with regularization) is minimized over all $K$ segments at once. As can be seen from the solution in Table~\ref{tab:mmse-bw}, this is equivalent to first estimating the autocorrelation matrix and cross-correlation vector via averaging the sample estimates per segment, whereafter one decoder is computed. It is easy to see that this is mathematically equivalent to concatenating all the data in one big matrix $\mat{X} \in \R^{KT \times LC}$ and vector $\vec{s}_{\textnormal{a}} \in \R^{KT \times 1}$ and computing the decoder straightforwardly. As such, it is an example of the \emph{early integration} paradigm, versus \emph{late integration} in the former case when averaging $K$ separate decoders. Both versions are included in our comparative study.
	
	Table~\ref{tab:mmse-bw} shows the four different flavors of the MMSE/LS-based decoder that were proposed as different AAD algorithms in~\cite{o2014attentional,biesmans2017auditory,Alick2019}, adopting different regularization techniques (L\textsubscript{$2$}/L\textsubscript{$1$}-regularization) or ways to train the decoder (averaging decoders or correlation matrices).
	
	\begin{table*}
		\footnotesize
		\centering
		\begin{tabular}{p{3.4cm}p{5.5cm}p{5.09cm}}
			\toprule
			\textbf{Method} & \textbf{Cost function} & \textbf{Solution} \\
			\midrule
			Ridge regression + \newline averaging of decoders~\cite{o2014attentional} \newline(MMSE-avgdec-ridge) & $\hat{\vec{d}}_k = \underset{\vec{d}}{\textnormal{argmin}} \norm{\vec{s}_{\textnormal{a}_k}-\mat{X}_k\vec{d}}_2^2 + \lambda z_k \norm{\vec{d}}_2^2$
			& $\hat{\vec{d}}_k = \inverse{\left( \transpose{\mat{X}}_k\mat{X}_k + \lambda z_k\mat{I}\right)} \transpose{\mat{X}}_k\vec{s}_{\textnormal{a}_k}$ \newline and $\hat{\vec{d}} = \frac{1}{K} \sumlim{k = 1}{K}\hat{\vec{d}}_k$\\
			Lasso + \newline averaging of decoders~\cite{Alick2019} \newline(MMSE-avgdec-lasso) & $\hat{\vec{d}}_k = \underset{\vec{d}}{\textnormal{argmin}} \norm{\vec{s}_{\textnormal{a}_k}-\mat{X}_k\vec{d}}_2^2 + \lambda q_k \norm{\vec{d}}_1$ & ADMM and $\hat{\vec{d}} = \frac{1}{K} \sumlim{k = 1}{K}\hat{\vec{d}}_k$ \\
			Ridge regression + \newline averaging of correlation matrices~\cite{biesmans2017auditory} \newline (MMSE-avgcorr-ridge) & $\hat{\vec{d}} = \underset{\vec{d}}{\textnormal{argmin}} \sumlim{k = 1}{K}\norm{\vec{s}_{\textnormal{a}_k}-\mat{X}_k\vec{d}}_2^2 + \lambda z \norm{\vec{d}}_2^2$ & $ \hat{\vec{d}} = \inverse{\left( \sumlim{k = 1}{K}\transpose{\mat{X}}_k\mat{X}_k + \lambda z\mat{I}\right)}\!\!\sumlim{k = 1}{K}\transpose{\mat{X}}_k\vec{s}_{\textnormal{a}_k}$ \\
			Lasso + averaging of correlation matrices~\cite{Alick2019} \newline(MMSE-avgcorr-lasso) & $\hat{\vec{d}} = \underset{\vec{d}}{\textnormal{argmin}} \sumlim{k = 1}{K}\norm{\vec{s}_{\textnormal{a}_k}-\mat{X}_k\vec{d}}_2^2 + \lambda q \norm{\vec{d}}_1$ & ADMM \\
			\hline
		\end{tabular}
		\caption{A summary of the supervised backward MMSE-decoder and its different flavors.}
		\label{tab:mmse-bw}
	\end{table*}
	\noindent
	
	\subsubsection{Canonical correlation analysis (CCA)}
	\label{sec:cca}
	\noindent
	CCA to decode the auditory brain has been proposed in~\cite{Dmochowski2018,de2018decoding}. It has been applied to the AAD problem for the first time in~\cite{Alick2019}. CCA combines a spatio-temporal backward (decoding) model $\vec{w}_{x} \in \R^{LC \times 1}$ on the EEG and a temporal forward (encoding) model $\vec{w}_{s_\textnormal{a}} \in \R^{L_{\textnormal{a}} \times 1}$ on the speech envelope, with $L_{\textnormal{a}}$ the number of filter taps of the encoding filter. In this sense, CCA differs from the previous approaches, which were all different flavors of the same MMSE/LS-based decoder. In CCA, both the forward and backward model are estimated \emph{jointly} such that their outputs are maximally correlated:
	\begin{equation}
		\label{eq:cca}
		\underset{\vec{w}_{x},\vec{w}_{s_\textnormal{a}}}{\textnormal{max}}  \frac{\ev{\left(\transpose{\vec{w}}_{x}\vec{x}(t)\right)\left(\transpose{\vec{w}}_{s_\textnormal{a}}\vec{s}_{\textnormal{a}}(t)\right)}}{\sqrt{\ev{\left(\transpose{\vec{w}}_{x}\vec{x}(t)\right)^2}}\sqrt{\ev{\left(\transpose{\vec{w}}_{s_\textnormal{a}}\vec{s}_{\textnormal{a}}(t)\right)^2}}}
		= \underset{\vec{w}_{x},\vec{w}_{s_\textnormal{a}}}{\textnormal{max}}  \frac{\transpose{\vec{w}}_x\mat{R}_{xs_\textnormal{a}}\vec{w}_{s_\textnormal{a}}}{\sqrt{\transpose{\vec{w}}_x\mat{R}_{xx}\vec{w}_x}\sqrt{\transpose{\vec{w}}_{s_\textnormal{a}}\mat{R}_{s_\textnormal{a}s_\textnormal{a}}\vec{w}_{s_\textnormal{a}}}},
	\end{equation}
	\noindent
	where $\vec{s}_\textnormal{a}(t) = \transpose{\begin{bmatrix} s_\textnormal{a}(t) & s_\textnormal{a}(t-1) & \cdots & s_\textnormal{a}(t-L_\textnormal{a}+1) \end{bmatrix}} \in \R^{L_\textnormal{a} \times 1} $. As opposed to the EEG filter $\vec{w}_{x}$, the audio filter $\vec{w}_{s_\textnormal{a}}$ is a causal filter, as the stimulus precedes the brain response. The solution of the optimization problem in~\eqref{eq:cca} can be easily retrieved by solving a generalized eigenvalue decomposition (details in~\cite{biesmans2017auditory,Alick2019}).

	In CCA, the backward model $\vec{w}_x$ and forward model $\vec{w}_{s_\textnormal{a}}$ are extended to a set of $J$ filters $\mat{W}_x \in \R^{LC \times J}$ and $\mat{W}_{s_\textnormal{a}} \in \R^{L_\textnormal{a} \times J}$ for which the outputs are maximally correlated, but mutually uncorrelated (the $J$ outputs of $\transpose{\mat{W}}_x \vec{x}(t)$ are uncorrelated to each other and the $J$ outputs of $\transpose{\mat{W}}_{s_\textnormal{a}} \vec{s}_\textnormal{a}(t)$ are uncorrelated to each other). There are now thus $J$ Pearson correlation coefficients between the outputs of the $J$ backward and forward filters (aka canonical correlation coefficients), which are collected in the vector $\vec{\rho}_i \in \R^{J \times 1}$ for speaker $i$, whereas before, there was only one per speaker. Furthermore, because of the way CCA constructs the filters, it can be expected that the first components are more important than the later ones. To find the optimal way of combining the canonical correlation coefficients, a linear discriminant analysis (LDA) classifier can be trained, as proposed in~\cite{de2018decoding}. To generalize the maximization of the correlation coefficients of the previous AAD algorithms (which is equivalent to taking the sign of the difference of the correlation coefficients of both speakers), we propose here to construct a feature vector $\vec{f} \in \R^{J \times 1}$ by subtracting the canonical correlation vectors: $\vec{f} = \vec{\rho}_1 - \vec{\rho}_2$, and classify $\vec{f}$ with an LDA classifier. As proposed in~\cite{de2018decoding}, we use PCA as a preprocessing step on the EEG to reduce the number of parameters. In fact, this is a way of regularizing CCA and can as such be viewed as an alternative to the regularization techniques proposed in other methods.
	
	\subsubsection{Training-free MMSE-based with lasso (MMSE-adap-lasso)}
	\label{sec:miran}
	\noindent
	In~\cite{miran2018real}, a fundamentally different AAD algorithm is proposed. In this comparative study, all other AAD algorithms are \emph{supervised}, batch-trained algorithms, which have a separate training and testing stage. First, the decoders need to be trained in a supervised manner using a large amount of ground-truth data, after which they can be applied to new test data. In practice, this necessitates a (potentially cumbersome) a priori training stage, resulting in a fixed decoder, which does not adapt to the non-stationary EEG signal characteristics, e.g., due to changing conditions or brain processes. The AAD algorithm in~\cite{miran2018real} aims to overcome these issues by adaptively estimating a decoder for each speaker and simultaneously using the outputs to decode the auditory attention. Therefore, this training-free AAD algorithm has the advantage of adapting the decoders to non-stationary signal characteristics, however, without requiring the same large amount of ground-truth data as the supervised AAD algorithms.
	
	In this comparative study, we have removed the state-space and dynamic decoder estimation modules to produce a single decision for each decision window, similar to the other AAD algorithms in this study (the full description of the algorithm can be found in~\cite{miran2018real}). This leads to the following formulation:
	\begin{equation}
		\label{eq:ls-adap-lasso}
		\begin{aligned}
			&\hat{\vec{d}}_{i,l} = \underset{\vec{d}}{\textnormal{argmin}} \norm{\vec{s}_{i,l}-\mat{X}_l\vec{d}}_2^2 + \lambda q \norm{\vec{d}}_1,
		\end{aligned}
	\end{equation}
	for the $i$\textsuperscript{th} speaker in the $l$\textsuperscript{th} decision window. In the context of AAD, for every new incoming window of $\tau$ seconds of EEG and audio data, two decoders are thus estimated (one for each speaker). As an attentional marker, these estimated decoders could be applied to the EEG data $\mat{X}_l$ of the $l$\textsuperscript{th} decision window to compute the correlation with their corresponding stimuli envelopes. In addition, the authors of~\cite{miran2018real} propose to identify the attended speaker by selecting the speaker with the largest L\textsubscript{$1$}-norm of its corresponding decoder $\hat{\vec{d}}_{i,l}$, as the attended decoder should exhibit more sparse, significant peaks, while the unattended decoder should have smaller, randomly distributed coefficients. The regularization parameter is again being cross-validated and defined in the same way as for MMSE-avgdec/corr-lasso. To prevent overfitting by decreasing the number of parameters to be estimated, the authors of~\cite{miran2018real} have proposed to a priori select a subset of EEG channels. In our comparative study, we also adopt this approach and select the same channels.
	
	While we do not adopt the extra post-processing state-space modeling steps from~\cite{miran2018real,aroudi2020improving} in order to focus on the core AAD algorithm, it is noted that such an extra smoothing step, which also takes previous and/or future decisions into account, can effectively enhance the performance of most AAD algorithms, albeit at the cost of a potential algorithmic delay in the detection of attention switches~\cite{miran2018real}.
	
	\subsection{Nonlinear methods}
	\label{sec:nonlinear-methods}
\begin{figure}
		\centering
		\begin{subfigure}{1\linewidth}
			\centering
			\includegraphics[width=1\linewidth]{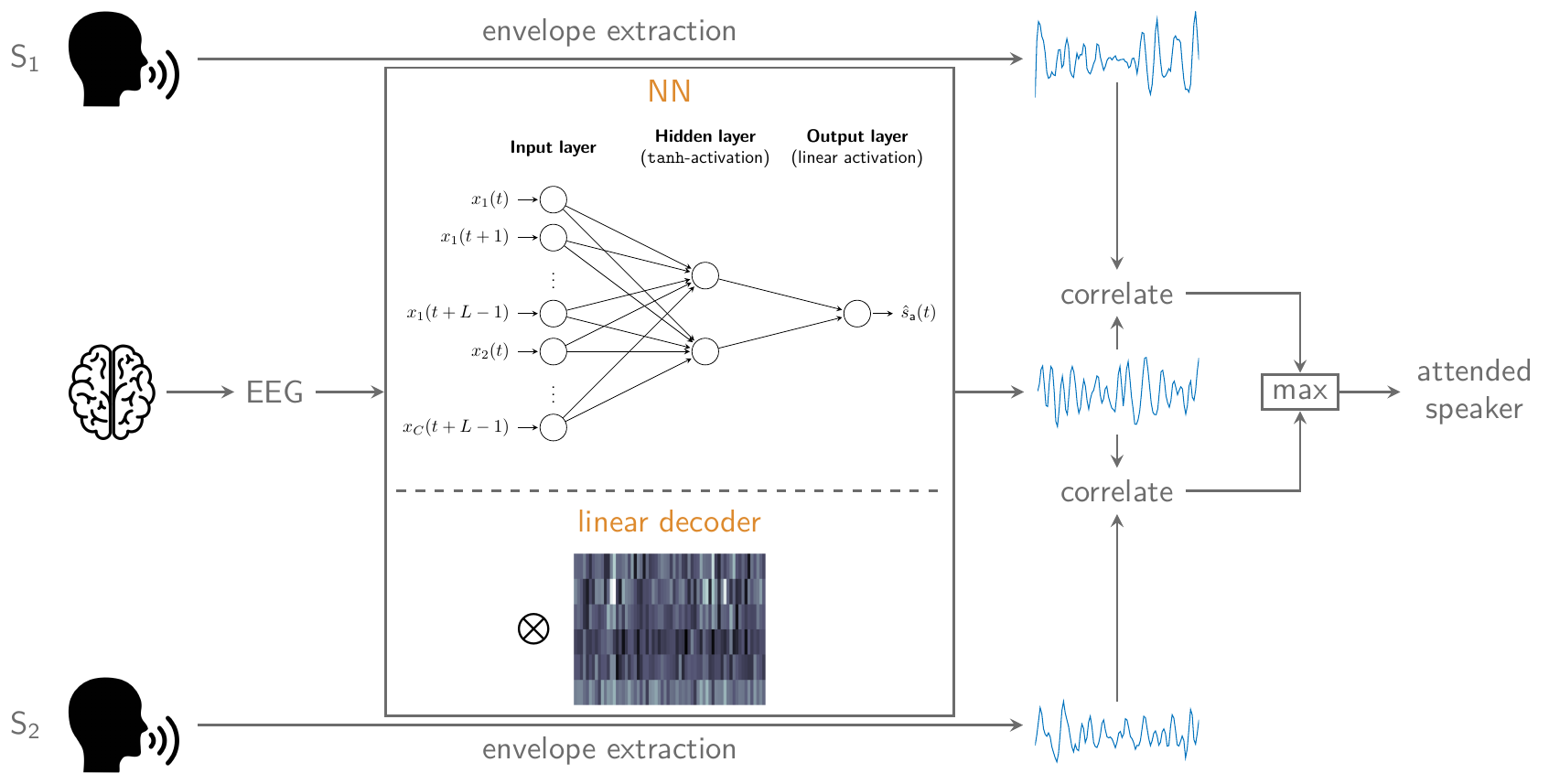}
			\caption{}
			\label{fig:overview-stim-rec-nn-sr}
		\end{subfigure}
		
		\begin{subfigure}{1\linewidth}
			\centering
			\includegraphics[width=1\linewidth]{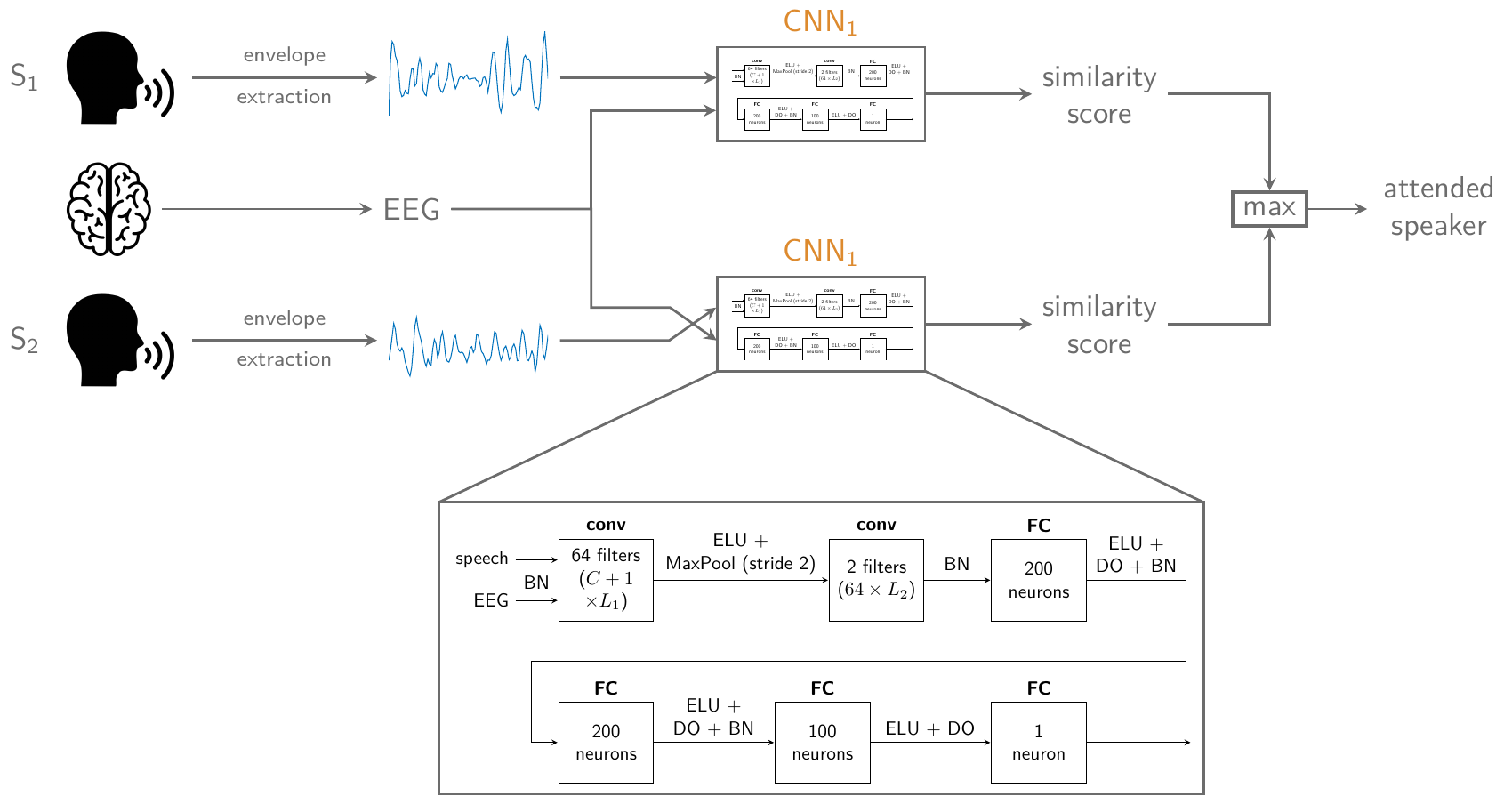}
			\caption{}
			\label{fig:overview-cnn-sim}
		\end{subfigure}
		
		\begin{subfigure}{1\linewidth}
			\centering
			\includegraphics[width=0.93\linewidth]{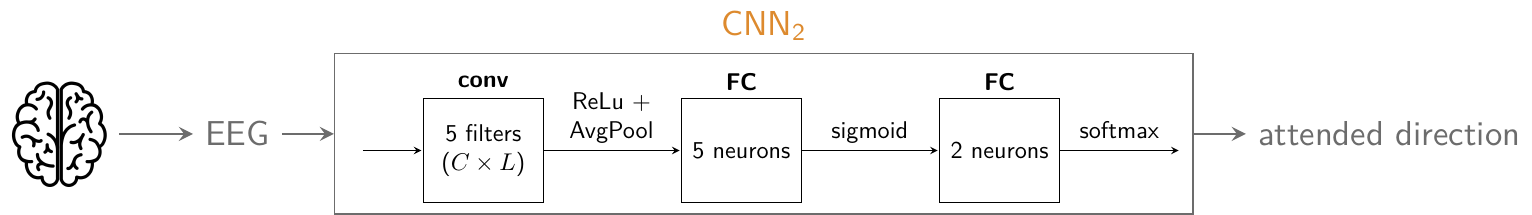}
			\caption{}
			\label{fig:overview-cnn-loc}
		\end{subfigure}
		\caption{A conceptual overview of the different AAD algorithms and the different network topologies of (a) a linear stimulus reconstruction decoder and NN-SR, (b) CNN-sim, and (c) CNN-loc. `conv' = convolutional layer, `FC' = fully connected layer, `BN' = batch normalization, `ELU' = exponential linear unit, `ReLu' = rectified linear unit, `DO' = drop-out, `MaxPool' = max-pooling, and `AvgPool' = average pooling.}
		\label{fig:concp-overview-algos}
	\end{figure}

	\noindent
	Nonlinear methods based on (deep) neural networks can adopt a stimulus reconstruction approach~\cite{de2017machine} similar to the linear methods, but can also classify the attended speaker directly from the EEG and the audio (aka direct classification)~\cite{Vandecappelle475673,Ciccarelli2019}. However, these nonlinear methods are more vulnerable to overfitting~\cite{Vandecappelle475673}, in particular for the small-size datasets that are typically collected in AAD research. In order to appreciate the differences between current neural network-based AAD approaches, Fig.~\ref{fig:concp-overview-algos} shows a conceptual overview of the different strategies and network topologies of the presented nonlinear methods. We give a concise description of each architecture below, but refer to the respective papers for further details. 
	
	\subsubsection{Fully connected stimulus reconstruction neural network (NN-SR)}
	\label{sec:nn-sr}
	\noindent
	In~\cite{de2017machine}, the authors proposed a fully-connected neural network with a single hidden layer that reconstructs the envelope based on a segment of EEG. As shown in Fig.~\ref{fig:overview-stim-rec-nn-sr}, the input layer consists of $LC$ neurons (similar to a linear decoder), with $L$ the number of time lags and $C$ the number of EEG channels. These neurons are connected to a hidden layer with two neurons and a $\tanh$ activation function. These two neurons are then finally combined into a single output neuron that uses a linear activation function and outputs one sample of the reconstructed envelope. As such, the network has $2\times (LC+1)$ (hidden layer) $+2+1$ (output layer) $\approx 3446$ trainable parameters.
	
	The network is trained to minimize $1-\rho\!\left(\hat{\vec{s}}_{\text{a}},\vec{s}_{\text{a}}\right)$ over a segment of $M$ training samples (within this segment the neural network coefficients are kept constant), with $\rho\!\left(\cdot\right)$ the Pearson correlation coefficient, and $\hat{\vec{s}}_{\text{a}}, \vec{s}_{\text{a}} \in \R^{M \times 1}$ the reconstructed and attended envelope, respectively. Minimizing this cost function is equivalent to maximizing the Pearson correlation coefficient between the reconstructed and attended speech envelope, similar to linear stimulus reconstruction approaches. The trained network is then used as a decoder, where the speech envelope showing the highest correlation with the decoder output is selected as the attended speaker. This algorithm can be extended to more than two competing speakers similar to the other linear stimulus reconstruction decoders.
	
	\subsubsection{Convolutional neural network to compute similarity between EEG and stimulus (CNN-sim)}
	\label{sec:cnn-c}
	\noindent
	In~\cite{Ciccarelli2019}, a convolutional neural network (CNN) is proposed to directly compare a $C \times T$ EEG segment with a $1 \times T$ speech envelope. This network is trained to output a similarity score $\in [0,1]$ (similar to the correlation coefficient used in other approaches) between the EEG and the speech envelope using a binary cross-entropy cost function. The speech envelope that, according to the trained CNN, is most similar to the EEG is then identified as the attended speaker. This approach can be easily extended to more than two speakers by computing a similarity score for each speaker, and taking the maximum over all scores to identify the attended speaker.
	
	The network, depicted in Fig.~\ref{fig:overview-cnn-sim}, consists of two convolutional layers, with max-pooling (stride two) after the first convolutional layer, and four fully connected (FC) layers. In total, this network has $64 \times (C+1) \times L_1$ (first convolutional layer) $+ 2 \times 64 \times L_2$ (second convolutional layer) $+ 200 \times 3$ (first fully connected layer) $+ 200 \times 201$ (second FC layer) $+ 100\times 201$ (third FC layer) $+ 101$ (fourth FC layer) $\approx 69070$ trainable parameters. An exponential linear unit is used as a nonlinear activation function. Furthermore, drop-out is used as a regularization technique to prevent overfitting in the FC layers, while also batch normalization is used throughout the network. Details about the training can be found in~\cite{Ciccarelli2019}.
	
	\subsubsection{Convolutional neural network to determine spatial locus of attention (CNN-loc)}
	\label{sec:cnn-s}
	\noindent
	In~\cite{Vandecappelle475673}, a CNN is proposed to determine the spatial locus of attention (i.e., the directional focus of attention, e.g., left or right), solely based on the EEG. This is a fundamentally different approach to tackle the AAD problem, which has the advantage of not requiring the individual speech envelopes (see also Section~\ref{sec:sp-sep-eeg-minia}). Furthermore, it avoids the requirement to estimate a correlation coefficient over a relatively long decision window length as in all aforementioned algorithms, thereby avoiding large algorithmic delays.
	
	This CNN determines the spatial locus of attention, starting from a $C \times T$ EEG segment. As shown in Fig.~\ref{fig:overview-cnn-loc}, it consists of one convolutional layer and two fully connected layers. The convolutional layer consists of five spatio-temporal filters, with lags $L$ similar to before, each outputting a one-dimensional time series of length $T$, on which a rectified linear unit activation function is applied. Afterwards, an average pooling layer condenses each output series into a scalar, leading to a five-dimensional vector. This vector is then used as an input for two fully connected layers, the first one consisting of five neurons with a sigmoid activation function and the output layer consisting of two neurons and a softmax layer. In total, this network has $5\times C \times L$ (convolutional layer) $+ 5 \times 6$ (first FC layer) $+2 \times 6$ (second FC layer) $\approx 2708$ trainable parameters. The CNN can be extended to more than two possible spatial locations (and thus competing speakers) by adding more output neurons to the network to generalize it to a multi-class problem, in which each class corresponds to a location or zone in which the attended speaker is believed to be positioned.
	
	A cross-entropy cost function is minimized using mini-batch gradient descent. Weight decay regularization is applied, as well as a post-training selection of the optimal model based on the validation loss. Furthermore, during training, not only data from the subject under test (as in all other methods) but also data from other subjects are used, as it was found in~\cite{Vandecappelle475673} that this prevents the model from overfitting on the training data in case only a limited amount of data of the subject under test is available. Therefore, this inclusion of data from other subjects can be seen as a type of regularization.
	
	\section{Comparative study of AAD algorithms}
	\label{sec:comp-study}
	\noindent
	We compared the aforementioned AAD algorithms on two publicly available datasets~\cite{das_neetha_2019_3377911,fuglsang_soren_a_2018_1199011} in a subject-specific manner. Both datasets have been collected with the purpose of AAD, using a competing talker setup in which two stories are simultaneously narrated. Details on the datasets and the preprocessing of the EEG and audio data are described in \textbf{[Pop-out box 1]}. All algorithms, including the deep learning methods, are re-trained from scratch on each dataset separately.
	
	Given a decision window length $\tau$, the performance of each algorithm is evaluated via the accuracy $p\in[0,100]\%$, defined as the percentage of correctly classified decision windows. Since EEG is the superimposed activity of many different (neural) processes, the correlation $\rho$ between the reconstructed and attended envelope is typically quite low (in the order of $0.05$-$0.2$). Therefore, it is important to use a sufficiently long decision window such that the decision process is less affected by estimation noise in $\rho$ due to the finite sample size. As a result, the accuracy $p$ generally increases for longer decision window lengths $\tau$, leading to a so-called `$p(\tau)$-performance curve'. These accuracies are obtained using the cross-validation procedure described in \textbf{[Pop-outbox 2]}.
	
	This $p(\tau)$-performance curve thus presents a trade-off between accuracy and decision delay of the AAD system (a long decision length implies a slower reaction time to a switch in attention). In~\cite{geirnaert2020interpretable}, the \emph{minimal expected switch duration} (MESD) metric has been proposed to resolve this trade-off in order to compare AAD algorithms more easily. The MESD metric determines the most optimal point on the $p(\tau)$-performance curve in the context of attention-steered gain control by minimizing the expected time it takes to switch the gain between two speakers in an optimized robust gain control system. As such, it outputs a single-number time metric (the MESD [s]) for a $p(\tau)$-performance curve and thus removes the loss of statistical power due to multiple-comparison corrections in statistical hypothesis testing (due to testing for multiple decision window lengths). Furthermore, the MESD ensures that the statistical comparison is automatically focused on the most practically relevant points on the $p(\tau)$-performance curve, which typically turn out to be the ones corresponding to short decision window lengths $\tau<\SI{10}{\second}$~\cite{geirnaert2020interpretable}. A higher MESD corresponds to a worse AAD performance and vice versa. This MESD metric is a theoretical metric that is not based on actual attention switches in the data, which are also not present in the datasets used. It is merely used here as a comparative metric, which does not necessarily reflect the true switching time as it relies on independence assumptions in the underlying Markov model, which can be violated in practice.
	
	\subsection{Statistical analysis}
	\label{sec:stat-analysis}
	\noindent
	To statistically compare the included AAD algorithms, we adopt a linear mixed-effects model (LMM) on the MESD values with the AAD algorithm as a fixed effect and with subjects as a repeated-measure random effect. Five contrasts of interest were set a priori according to the binary tree structure in Fig.~\ref{fig:tree-algos}. Algorithms that were not competitive or did not perform significantly better than chance are excluded from the statistical analysis, which is why some algorithms are not included in the contrasts (see Section~\ref{sec:perf-curves}). The planned contrasts reflect the most important different features between AAD algorithms, as shown in Fig.~\ref{fig:tree-algos}, motivating how they are set. The significance level is set at $\alpha = 0.05$.
	
	\subsection{Results}
	\label{sec:res-disc}
	
	\subsubsection{Performance curves}
	\label{sec:perf-curves}
\begin{figure}
		\centering
		\begin{subfigure}{0.5\linewidth}
			\centering
			\includegraphics[width=1\linewidth]{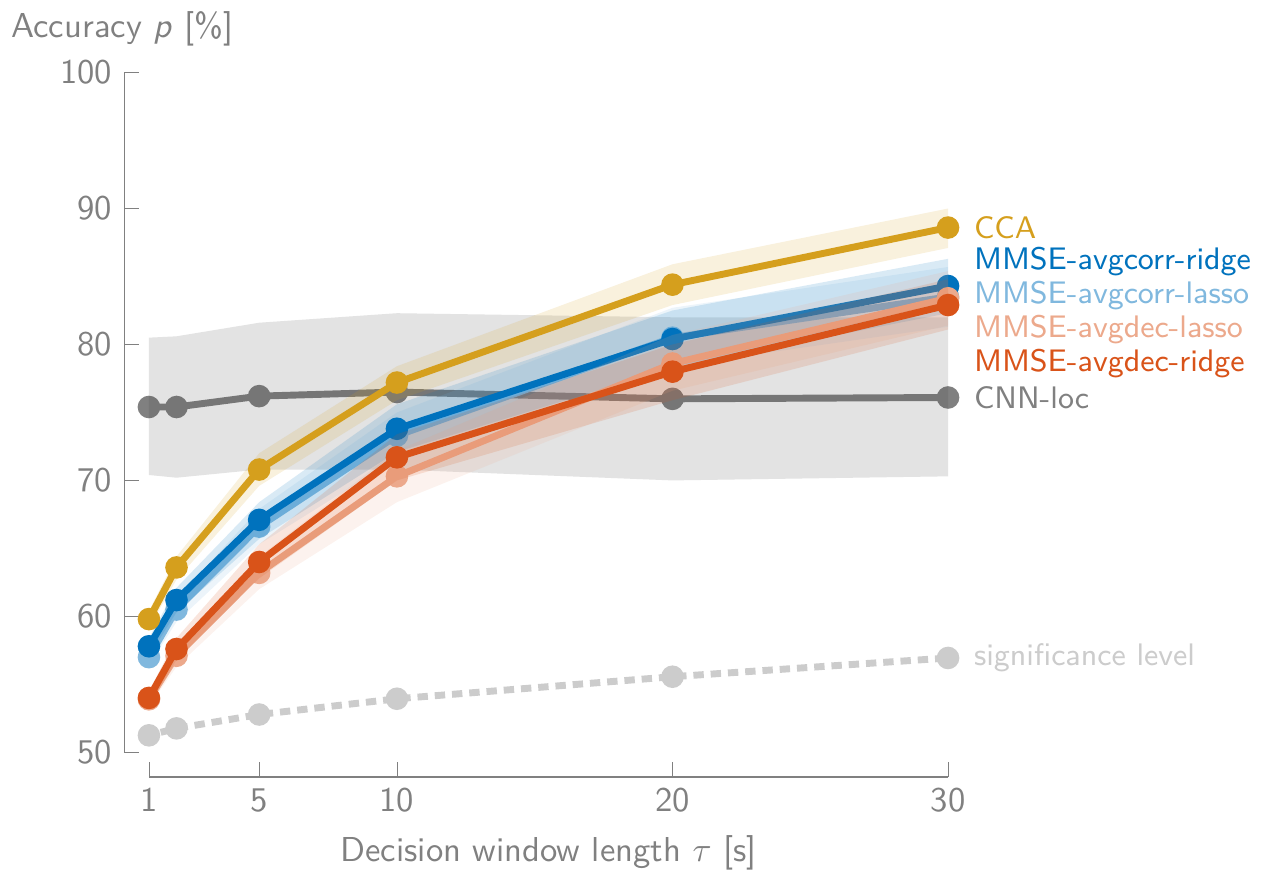}
			\caption{}
			\label{fig:perf-curve-das}
		\end{subfigure}%
		\begin{subfigure}{0.5\linewidth}
			\centering
			\includegraphics[width=1\linewidth]{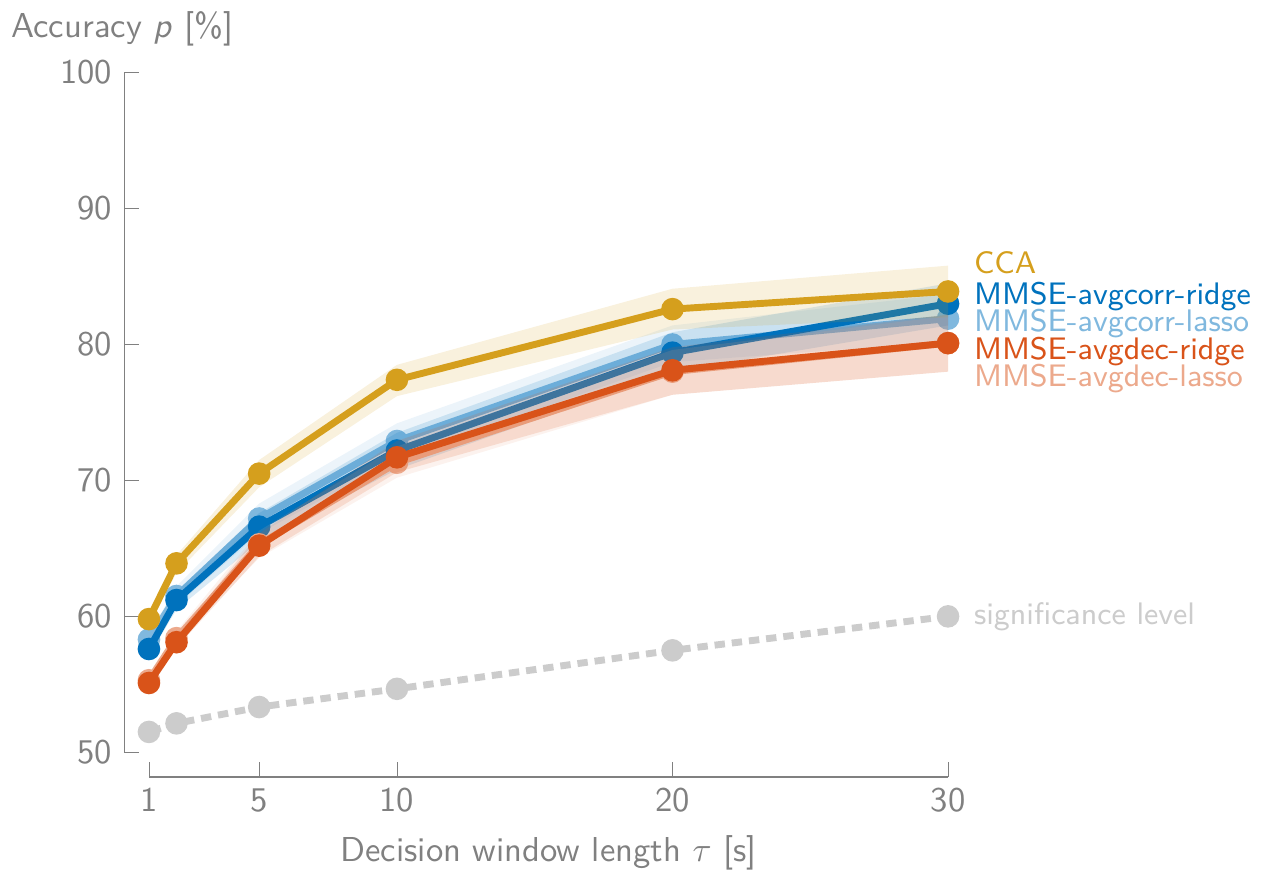}
			\caption{}
			\label{fig:perf-curve-fuglsang}
		\end{subfigure}
		\caption{The accuracy $p$ (mean $\pm$ standard error on the mean across subjects) as a function of the decision window length $\tau$ for (a) Das-2015 and (b) Fuglsang-2018. MMSE-adap-lasso, CNN-sim, and NN-SR did not perform significantly better than a random classifier and are not depicted. CNN-loc achieved competitive results only on the Das-2015 dataset. }
		\label{fig:perf-curves}
	\end{figure}

	\noindent
	Fig.~\ref{fig:perf-curves} shows the $p(\tau)$-performance curves of the different AAD algorithms on both datasets. For the MMSE-based decoders, it is observed that there is barely an effect of the type of regularization and that averaging correlation matrices (early integration) consistently outperforms averaging decoders (late integration). Furthermore, CCA outperforms all other linear algorithms. Lastly, on Das-2015, it is clear that decoding the spatial locus of attention using \mbox{CNN-loc} substantially outperforms the stimulus reconstruction methods for short decision windows ($< \SI{10}{\second}$), where CNN-loc appears to be less affected by the decision window length. However, the standard error on the mean is much higher for the CNN-loc algorithm than for the other methods, indicating a higher inter-subject variability.
	
	The performances of MMSE-adap-lasso, CNN-sim, and NN-SR are not shown in Fig.~\ref{fig:perf-curves} as they did not exceed the significance level or were not competitive on either of the two datasets. For a decision window length of $\SI{10}{\second}$, the MMSE-adap-lasso algorithm achieves an average accuracy of $52.9\%$ with a standard deviation of $4.3\%$ on the Das-2015 dataset and $49.8\%$ with a standard deviation of $5.9\%$ on the Fuglsang-2018 dataset. The CNN-sim algorithm achieves $51.7\%$ on average with a standard deviation of $2.3\%$ on the Das-2015 dataset (where there was no convergence for 5 subjects) and $58.1\%$ with a standard deviation of $9.2\%$ on the Fuglsang-2018 dataset. Lastly, the NN-SR algorithm achieves on average only $52.1\%$ (standard deviation $4.4\%$) on the Das-2015 dataset and $52.3\%$ (standard deviation $3.6\%$) on the Fuglsang-2018 dataset. As these algorithms did not significantly outperform a random classifier or were not competitive, they were also excluded from the statistical analysis. Furthermore, CNN-loc did not perform well on Fuglsang-2018 (i.e., $56.3\%$ with a standard deviation of $4.5\%$ on $\SI{10}{\second}$ decision windows). As such, planned contrast I was also excluded from the analysis for the Fuglsang-2018 dataset.
	
	\subsubsection{Subject-specific MESD performance}
	\label{sec:mesd-perf}
\begin{figure}
		\centering
		\begin{subfigure}{0.5\linewidth}
			\centering
			\includegraphics[width=1\linewidth]{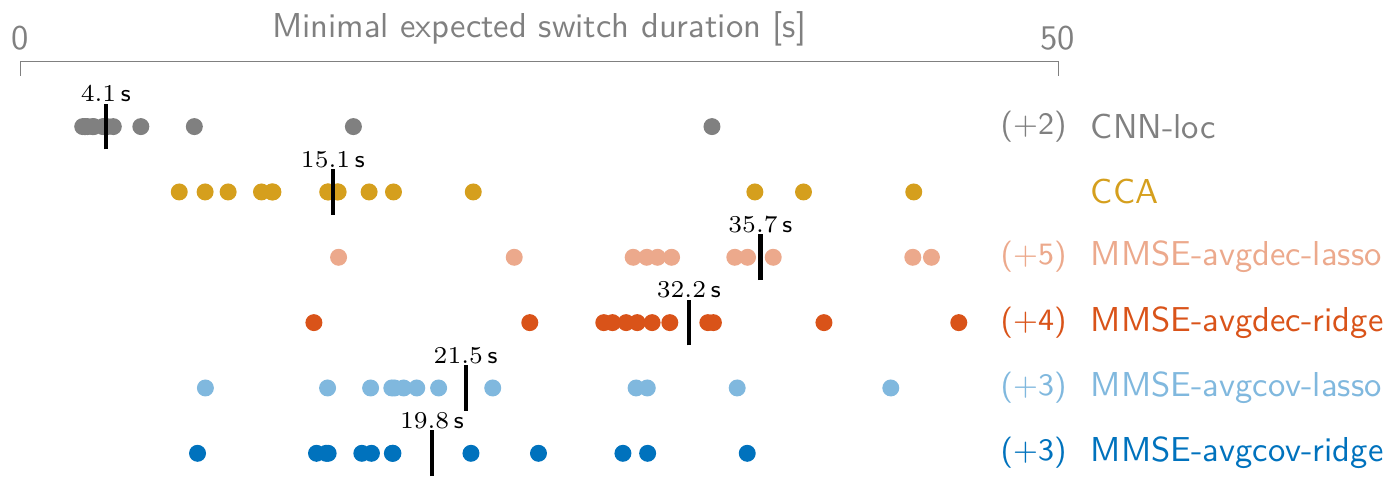}
			\caption{}
			\label{fig:mesd-das}
		\end{subfigure}%
		\begin{subfigure}{0.5\linewidth}
			\centering
			\includegraphics[width=1\linewidth]{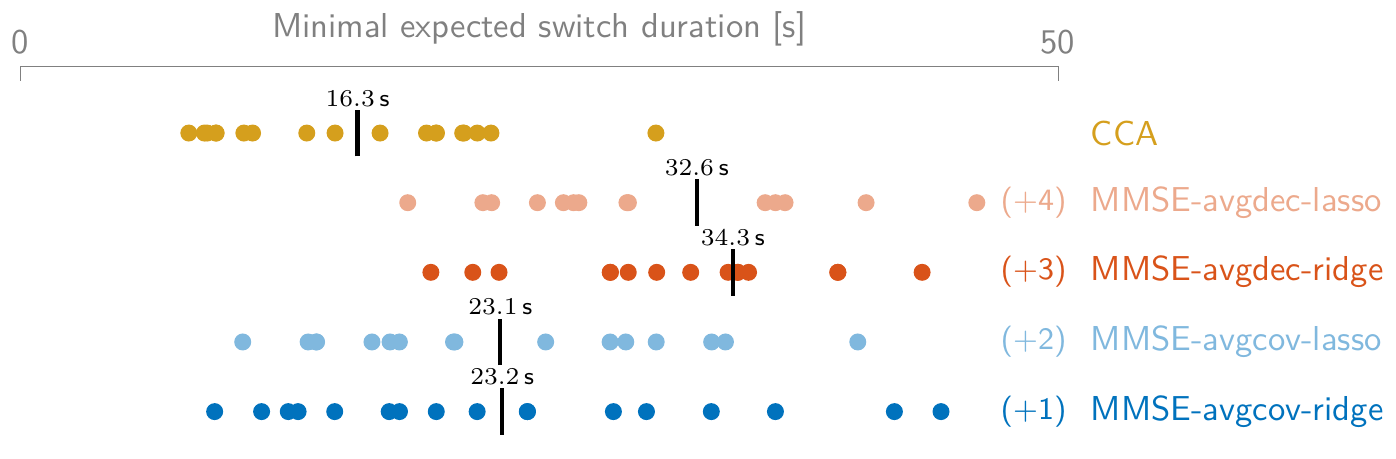}
			\caption{}
			\label{fig:mesd-fuglsang}
		\end{subfigure}
		\caption{The per subject MESD values, with the median indicated with a bar, for (a) Das-2015 and (b) Fuglsang-2018. The number of data points with an MESD $>\SI{50}{\second}$ are indicated as (+x). However, these were included in the computation of the medians.}
		\label{fig:mesd}
	\end{figure}

	\noindent
	A visual analysis of the per-subject MESD values (Fig.~\ref{fig:mesd}) confirms the trends based on the performance curves. These trends are also confirmed by the statistical analysis\footnote{The two outlying subjects of the CNN-loc algorithm were removed in all comparisons on the Das-2015 dataset.} using the LMM. There indeed is a significant improvement when decoding the spatial locus of attention via a nonlinear method versus the linear stimulus reconstruction methods ($p <0.001$ (Das-2015)). Furthermore, CCA significantly outperforms all backward stimulus reconstruction decoders ($p < 0.001$ (Das-2015), $p < 0.001$ (Fuglsang-2018)), while there is also a significant improvement when averaging correlation matrices compared to averaging decoders ($p = 0.0028$ (Das-2015), $p < 0.001$ (Fuglsang-2018)). There is no significant effect of the specific regularization technique ($p = 0.79$ (Das-2015), $p = 0.30$ (Fuglsang-2018) in averaging correlation matrices; $p = 0.57$ (Das-2015), $p = 0.91$ (Fuglsang-2018) in averaging decoders).
	
	\subsection{Discussion}
	\label{sec:perf-conclusions}
	\noindent
	From the results and statistical analysis, it is clear that CCA~\cite{de2018decoding}, which adopts a joint forward and backward model, outperforms the other stimulus reconstruction methods. Furthermore, the CNN-loc method~\cite{Vandecappelle475673}, which decodes the spatial locus of attention based on the EEG alone (i.e., without using the speech stimuli), substantially outperforms all stimulus reconstruction methods on the Das-2015 dataset at short decision window lengths, leading to substantially lower MESDs. This relatively high performance at short decision windows is attributed to the fact that this method avoids correlating the decoded EEG with the speech envelope, thereby not suffering from the noise-susceptible correlation estimation. However, the non-significant performance of CNN-loc on the Fuglsang-2018 dataset implies that alternative algorithms for decoding the spatial locus of attention might be required to improve robustness and generalization to different conditions. 
	\par
	Remarkably, while the traditional linear stimulus reconstruction methods are found to perform well across datasets, none of the tested nonlinear (neural network) methods achieve a competitive performance on \textit{both} benchmark datasets, even though high performances were obtained on the respective datasets used in~\cite{de2017machine,Vandecappelle475673,Ciccarelli2019}. This shows that these architectures do not always generalize well, even after re-training them on a new dataset (the original authors validated the implementations in our benchmark study to rule out potential discrepancies in the implementation). Due to the black-box nature of these methods, it remains unclear what causes success on one dataset and failure on another. One possible explanation is that the design process that eventually led to the reported network architecture was too tailored to a particular dataset (and its size), despite proper cross-validation. Furthermore, (deep) neural networks may potentially pick up subtle patterns that may change or become absent in different experimental set-ups due to differences in equipment, speech stimuli, or experiment protocols.
	
	Although this lack of reproducibility across datasets seems to undermine the practical usage of the presented nonlinear AAD methods, the current benchmark datasets are possibly too small for these methods to draw firm conclusions. AAD based on (deep) neural networks may become more robust when larger datasets become available, containing more subjects, more EEG data per subject, and more variation in experimental conditions. Nevertheless, the results of this comparative study point out the risks of overfitting and overdesigning these architectures, thereby emphasizing the importance of extensive validation with multiple independent datasets.
	
	\section{Open challenges and outlook}
	\label{sec:sp-sep-eeg-minia}
	\noindent
	
	\subsection{Validation in realistic listening scenarios}
	\label{sec:naturalistic-conditions}
	\noindent
	In this comparative study, we investigated and compared different AAD algorithms on data that have been collected in a very controlled environment, with only two competing speakers, without much background noise or heavy reverberation, with well-separated competing speakers, and without switches in attention. Many of these AAD algorithms still need to be further validated in more complex listening scenarios.
	
	While we tested the algorithms on data with only two competing speakers, the algorithm of~\cite{o2014attentional} has been extended to four competing speakers in~\cite{schafer2018testing} with a limited performance loss. Thus, it is hoped that all other flavors of this decoder, including the CCA and MMSE-adap-lasso extensions, and the NN-SR and CNN-sim models, which are based on the same principles, similarly generalize to multiple speakers. However, the effect of an increasing number of competing speakers and speaker locations beyond two for the CNN-loc algorithm is not immediately clear due to the fundamentally different decoding strategy. Decoding the spatial locus of attention may become much harder when there are more than two speaker locations. To what extent this affects the performance remains to be investigated.
	
	The impact of background noise (such as babble noise) and reverberation on the AAD performance for stimulus reconstruction decoders has been extensively investigated in~\cite{fuglsang_soren_a_2018_1199011,das2018eeg,aroudi2019impact}. For example, in~\cite{das2018eeg}, it was shown that the AAD accuracy even increases when there is moderate background noise compared to no noise. Similarly, in~\cite{aroudi2019impact}, the AAD performance was comparable across different noisy and reverberant conditions. Moreover, even when training decoders with data collected in different acoustic conditions (noise and reverberation) than the test condition, good AAD performance can be achieved. In~\cite{das2018eeg}, the effect of different speaker positions on the stimulus reconstruction decoder has been investigated as well, reporting better performance with increasing speaker separation, but also still acceptable accuracies for closely positioned competing speakers.
	
	Lastly, the effect of switches in auditory attention on the operation of several AAD algorithms is still unclear. While a theoretical analysis of the performance of AAD algorithms on attention switches has been performed in~\cite{geirnaert2020interpretable} and some preliminary results on artificial attention switches have been analyzed in~\cite{o2017neural}, the performance of AAD algorithms on natural attention switches largely remains to be investigated. 
	
	\subsection{Effects of speaker separation and denoising algorithms}
	\label{sec:sp-sep}
	\noindent
	As explained in Section~\ref{sec:review}, most AAD algorithms require access to the speech envelopes of the individual speakers. Although it is also possible to apply the stimulus reconstruction decoders for AAD on the unprocessed microphone signals, as shown in~\cite{van2017eeg,aroudi2019impact}, the performance then highly depends on a favorable relative position of the speakers and microphones. Thus, in the context of neuro-steered hearing devices, the extraction of the per-speaker envelopes from the hearing aid's microphone recordings is generally required. It is expected that the performed speaker separation is not perfect, affecting the quality of the speech envelopes, and thus also affecting the AAD algorithms that use these envelopes. Correspondingly, AAD algorithms that do not rely on this speaker separation step, such as decoding the spatial locus of attention~\cite{Vandecappelle475673}, have an inherent advantage. In any case, a speech enhancement algorithm is required to eventually extract the attended speaker, for which advanced and well-performing signal processing algorithms exist (e.g.,~\cite{luo2019conv}). 
	
	A few studies have already combined AAD with speaker separation and denoising algorithms, both using traditional beamforming approaches~\cite{van2017eeg,aroudi2020cognitive,pu2019joint,Das2020}, and deep neural networks for speaker separation~\cite{o2017neural,Han2019,Das2020}. Remarkably, many of these studies show only minor or hardly any effects on the AAD performance when using the demixed speech signals, even in challenging noisy conditions and despite significant distortions on the envelopes~\cite{aroudi2020cognitive,Das2020}. These positive results are paramount for the practical applicability of neuro-steered hearing devices.
	
	Finally, instead of treating the speaker extraction and AAD as two separate problems (as is the case in all aforementioned studies), one could also aim to solve both problems simultaneously. In~\cite{pu2019joint}, the speaker extraction and AAD problem are coupled together in a joint optimization problem, where the beamformer is enforced to generate an output signal that is correlated to the output of a backward MMSE neural decoder, showing promising results.
	
	\subsection{EEG miniaturization and wearability effects}
	\label{sec:miniaturization}
	\noindent
	The data used in this paper are recorded using expensive, heavy, bulky, and wet EEG recording systems. The realization of neuro-steered hearing devices requires a wearable, concealable EEG monitoring system. The research towards such concealable EEG systems is very active, resulting in novel miniature devices to acquire the EEG, for example, in the ear (e.g.,~\cite{Kappel2019}) or around the ear (e.g.,~\cite{Debener2015}). However, such wearable, concealable EEG systems, also called miniature EEG sensor devices, provide only a limited amount of EEG channels, which record brain activity within a small area. A first analysis using such an around-the-ear EEG system in the context of AAD showed potential, albeit with a significant decrease in performance~\cite{mirkovic2016target}.
	
	In another (top-down) approach, it was shown that using a data-driven selection of the best $10$ EEG channels of a standard $64$-channel EEG cap does not reduce the AAD performance for the linear stimulus reconstruction decoder~\cite{narayanan2019analysis}. Similarly, in~\cite{Ciccarelli2019}, the number of channels was reduced from $64$ to $18$ channels without any negative impact on the performance of their AAD system. Moreover, in~\cite{narayanan2019analysis}, it was also demonstrated that using EEG measured with strategically positioned electrode pairs with $< \SI{5}{\centi\meter}$ inter-electrode distance results in similar AAD performances as with standard long-distance montages. This is important for EEG miniaturization, where only a small number of electrodes within a small area are available per device.  
	
	As mentioned before, the data used in this paper are collected using a wet EEG system. Such a wet EEG system requires a trained professional to apply the electrode gel and to mount the system~\cite{kam2019systematic}. This seriously hampers the practical applicability. Alternatively, dry EEG systems, which are easier to apply and thus more user-friendly and more suitable for long-term recordings~\cite{kam2019systematic}, are being developed (e.g.,~\cite{Kappel2019}). Although~\cite{kam2019systematic} shows that these dry EEG systems can be used to record EEG with similar quality as the conventional wet EEG systems, and~\cite{Ciccarelli2019} briefly showed that a dry EEG system could achieve similar AAD performances, more extensive experimenting with dry EEG systems in the context of AAD is required, in particular in combination with miniaturization strategies~\cite{Kappel2019}.
	
	While these results indicate that AAD is possible with fewer EEG electrodes and with dry and/or miniaturized EEG systems, the development of unobtrusive and wearable EEG systems for AAD remains an important hurdle towards user-friendly and practical neuro-steered hearing devices.
	
	\subsection{Outlook}
	\label{sec:future}
	\noindent
	Several studies have demonstrated that it is possible to decode auditory attention from a non-invasive neurorecording technique such as EEG. In our comparative study, we have shown that most of these results are reproducible on different data sets. However, even for the best linear (stimulus reconstruction) method (CCA), the accuracy at short decision windows is still too low, potentially leading to too slow reactions of the system to shifts in auditory attention, as indicated by a median MESD of $\SI{15}{\second}$. The results of this study have demonstrated that an alternative strategy, such as decoding the spatial locus of attention, could significantly improve on these short decision window lengths. Although nonlinear (deep learning) methods are believed to be able to improve AAD performances substantially, our study has demonstrated that the reported results obtained by these methods are hard to replicate on multiple independent AAD datasets. A major future challenge for AAD research is the design of an algorithm or neural network architecture that reliably improves on short decision windows and which is reproducible on different independent datasets.
	
	Furthermore, most of the presented AAD algorithms require supervised training and are fixed during operation. To avoid cumbersome a priori training sessions for each individual user, as well as to adapt to the time-varying statistics of the EEG (e.g., in different listening scenarios), training-free or unsupervised adaptive AAD algorithms should be developed. While several steps have been made in that direction~\cite{miran2018real}, the results of this study show that we are still far away from a practical solution. Moreover, such online adaptive AAD algorithms are paramount in the development of closed-loop systems for neuro-steered hearing devices, in which the end-user can react to and interact with the AAD algorithm and speech enhancement system. The interplay between the algorithmic processes in the hearing device and the user could enable neurofeedback effects that significantly improve the performance of the hearing device~\cite{zink2017online}.
	
	Lastly, these AAD algorithms need to be further evaluated in real-life situations, taking various realistic listening scenarios into account, as well as on potential hearing device users~\cite{Fuglsang2020}. The individual building blocks of a neuro-steered hearing device (Fig.~\ref{fig:neuro-steered-HA}) need to be integrated, in which an AAD algorithm is combined with a reliable and low-latency speaker separation algorithm, a miniaturized EEG sensor system, and a smart gain control system.
	
	Despite the many challenges ahead, the application of neuro-steered hearing devices as a neurorehabilitative assistive device has shown to be within reach, having the potential to substantially improve the functionality and user-acceptance of future generations of hearing devices. 
	
	\section*{Pop-out boxes}
	\vspace{0.4cm}
	\begin{tcolorbox}[colback = white,breakable,left=5pt,right=5pt]
		\begin{center}
			\textbf{Pop-out box 1: Experiment details}
		\end{center}
		\vspace{-0.3cm}
		\noindent
		\textbf{Data:} The characteristics of both datasets are summarized in the following table:
		
		\begin{center}
			\scalebox{0.765}{
				\begin{tabular}{p{3.65cm}p{5.675cm}p{8.1cm}}
					\toprule
					\footnotesize\textbf{Attribute} &\footnotesize \textbf{Das-2015~\cite{das_neetha_2019_3377911}} &\footnotesize \textbf{Fuglsang-2018~\cite{fuglsang_soren_a_2018_1199011}} \\
					\midrule
					\footnotesize Number of subjects &\footnotesize $16$ &\footnotesize $18$ \\
					\footnotesize Amount of data (per subject) &\footnotesize $\SI{72}{\minute}$&\footnotesize $\SI{50}{\minute}$\\
					\footnotesize EEG system &\footnotesize 64-channel Biosemi (wet EEG) &\footnotesize 64-channel Biosemi (wet EEG)\\
					\footnotesize Speakers &\footnotesize male \& male &\footnotesize male \& female \\
					\footnotesize Azimuth direction sources &\footnotesize $\pm 90^\circ$ &\footnotesize $\pm 60^\circ$\\
					\footnotesize Acoustic room condition &\footnotesize dichotic and HRTF-filtered in anechoic room&\footnotesize  HRTF-filtered in anechoic, mildly, and highly reverberant room\\  
					\hline
			\end{tabular}}
		\end{center}
		\par
		\textbf{Speech envelope extraction:} The individual speech signals are passed through a gammatone filterbank, which roughly approximates the spectral decomposition as performed by the human auditory system. Per subband, the audio envelopes are extracted and their dynamic range is compressed using a powerlaw operation with exponent $0.6$, after which the subband envelopes are summed into a single broadband envelope~\cite{das2016effect}.
		\par
		\textbf{Frequency range:} For computational efficiency, the speech envelopes as well as the EEG signals are both downsampled to $f_s=\SI{64}{\hertz}$, and bandpass filtered between $\SIrange{1}{32}{\hertz}$~\cite{de2017machine,Ciccarelli2019,Vandecappelle475673}. For the linear algorithms, this was further reduced to $f_s=\SI{20}{\hertz}$ and $\SIrange{1}{9}{\hertz}$ in order to be able to reduce the number of parameters in the spatio-temporal decoders (linear stimulus reconstruction methods have been demonstrated not to exploit information above $\SI{9}{\hertz}$~\cite{das2016effect}).
		\par
		\textbf{Hyperparameter settings:} The decoder lengths and CNN kernel lengths are set as in the original papers. For all linear methods, this is $L = \SI{250}{\milli\second}$, for NN-SR $L = \SI{420}{\milli\second}$, for CNN-loc $L = \SI{130}{\milli\second}$, and for CNN-sim $L_1 = \SI{30}{\milli\second}$ (first layer) and $L_2 = \SI{10}{\milli\second}$ (second layer). For CCA, $\SI{1.25}{\second}$ is chosen as the encoder length. The full set of $64$ channels are used in all algorithms, except for MMSE-adap-lasso, where the same $28$ channels as in~\cite{miran2018real} are chosen to reduce the number of parameters (since the decoder is estimated on much less data). The regularization parameters are cross-validated using 10 values in the range $[10^{-6},0]$. For CCA, it turned out that retaining all PCA components for both datasets is optimal.
	\end{tcolorbox}
	
	\begin{tcolorbox}[colback = white,breakable,left=4pt,right=4pt]
		\begin{center}
			\textbf{Pop-out box 2: Details on cross-validation procedure}
		\end{center}
		\vspace{-0.3cm}
		\noindent
		\textbf{Two-stage cross-validation:} The different algorithms are evaluated via a two-stage cross-validation (CV) procedure applied per subject and decision window length. The AAD accuracy is determined via an outer leave-one-segment-out CV (LOSO-CV) loop. Per outer fold, the optimal hyperparameter is determined via an inner ten-fold CV loop on the training set of the outer loop. The length of each left-out segment in the outer loop is chosen equal to $\SI{60}{\second}$, which is split into smaller disjoint decision windows. For example, for a decision window length of $\SI{30}{\second}$, each left-out segment results in two decisions. Additional details per AAD algorithm are provided in the following table (standard CV corresponds to training on all but one segment, testing on the left-out segment):

		\begin{center}
			\scalebox{0.765}{
				\begin{tabular}{p{2.5cm}p{7.3cm}p{7.3cm}}
					\toprule
					\footnotesize\textbf{Method} & \footnotesize\textbf{Outer LOSO-CV loop} & \footnotesize\textbf{Inner 10-CV loop} \\
					\midrule
					\footnotesize MMSE-avgcorr-ridge/lasso & \footnotesize standard &\footnotesize optimization of $\lambda$ (independent of $\tau$, tuned based on largest value of $\tau$)\\\midrule
					\footnotesize MMSE-avgdec-ridge/lasso &\footnotesize training data of each fold is split into windows of the same size as $\tau$. A different decoder is estimated in each of these subwindows and the decoders are averaged across all training folds (similar to~\cite{o2014attentional}) &\footnotesize optimization of $\lambda$ (re-optimized for $\tau$ due to the dependency of the training procedure on $\tau$) \\\midrule
					\footnotesize CCA &\footnotesize standard, additional LOSO-CV loop to train and test LDA classifier &\footnotesize optimization of the number of canonical correlation coefficients $J$ as input for LDA (re-optimized for each $\tau$) \\\midrule
					\footnotesize MMSE-adap-lasso &\footnotesize optimization of $\lambda$ per $\tau$ and fold by taking hyperparameter with highest accuracy on training fold &\footnotesize / \\ \midrule
					\footnotesize NN-SR &\footnotesize standard &\footnotesize / \\ \midrule
					\footnotesize CNN-loc &\footnotesize LOSpO-CV instead of LOSO-CV, training \textit{and} testing redone for $\tau$ &\footnotesize / \\ \midrule
					\footnotesize CNN-sim &\footnotesize ten-fold CV instead of LOSO-CV (due to computation time), training \textit{and} testing redone for $\tau$ &\footnotesize / \\
					\hline
					\vspace{0cm}
			\end{tabular}}
		\end{center}

		\textbf{Leave-one-speaker-out CV:} When using the LOSO-CV method, the test set always contains a speaker that is also present in the training set. To avoid potential overfitting to speakers in the training set for the CNN-loc algorithm, we use the leave-one-speaker-out CV (LOSpO-CV) method for this algorithm, as proposed and explained in~\cite{Vandecappelle475673}. For the linear methods, there is no difference between the LOSO-CV and LOSpO-CV method. This is validated by performing $100$ runs per subject, with in each run another random CV split (using the same amount of folds as for LOSpO-CV). We then tested whether the LOSpO-CV performance significantly differs from the median of this empirical distribution (i.e., the median over all random splits) across all subjects. For the CCA method, which has most degrees of freedom to overfit, the difference between the LOSpO-CV and median random-CV accuracy is less than 1\% on 20s decision windows, and a paired Wilcoxon signed-rank test (over subjects) shows no significant difference ($W = 85, n = 16, p = 0.38$).

	\end{tcolorbox}
	
	\bibliographystyle{IEEEtran}
	\bibliography{biblio-abrv}
	
\end{document}